\documentclass[12pt]{aipproc}

\layoutstyle{6x9}

\newcommand{\be}{\begin{equation}}
\newcommand{\ee}{\end{equation}}
\newcommand{\bea}{\begin{eqnarray*}}
\newcommand{\eea}{\end{eqnarray*}}
\newcommand{\ba}{\begin{eqnarray}}
\newcommand{\ea}{\end{eqnarray}}
\newcommand{\bee}{\begin{enumerate}}
\newcommand{\ene}{\end{enumerate}}

\begin{document}

\title{ A Remarkable Relation in the Gauge Sector of
Electroweakdynamics}
\author{ Jean Pestieau}{address ={
Institut de Physique Th\'eorique, Universit\'e catholique de
Louvain,\\ Chemin du Cyclotron 2, B-1348 Louvain-la-Neuve,
Belgium \\ 
e-mail:pestieau@fyma.ucl.ac.be}}

\begin{abstract}
A precise empirical relation between the electromagnetic
coupling $\alpha (m_Z)$ and $\sin^2 \theta^\ell_{\mbox{eff}}$
---where $\theta^\ell_{\mbox{eff}}$ is the effective electroweak
mixing angle extracted from $Z$ leptonic decays--- is made
manifest:
$$
\alpha (m_Z) = {\sin^3 \theta^\ell_{\mbox{eff}} \ \cos
\theta^\ell_{\mbox{eff}}\over 4\pi}.$$
\end{abstract}

\maketitle 

  In this paper we propose an ansatz for the gauge couplings of the
Electroweak model which allows to derive numerical values for the mass
of the $Z$ boson and the leptonic weak mixing angle that are in remarkable
agreement with the corresponding experimental figures.

Consider $g(m_Z)$ and $g'(m_Z)$, the two running gauge couplings
corresponding to $SU(2)$ and $U(1)$, respectively, in
electroweakdynamics (Glashow-Weinberg-Salam model). They are defined
at the $Z$ mass scale and are dimensionless just as $e(m_Z) =
\sqrt{4\pi \alpha (m_Z)}$, where $e(m_Z)$ is the electromagnetic
coupling defined at the $Z$ mass scale. We have the well-known
rectangular triangle relations of the $SU(2) \times U(1)$ gauge
theory:
$$
g^2 (m_Z) + g'^2 (m_Z) = g^2_Z (m_Z),
\eqno{\mbox{(1a)}}
$$
$$
{1\over g^2 (m_Z)} + {1\over g'^2 (m_Z)} = {1\over e^2 (m_Z)},
\eqno{\mbox{(1b)}}
$$
$$
g(m_Z) \cdot g' (m_Z) = g_Z (m_Z) \cdot e (m_Z),
\eqno{\mbox{(1c)}}
$$
with $g_Z (m_Z)$, the $Z$ coupling.

The gauge couplings $g(m_Z), g' (m_Z)$ and $g_Z (m_Z)$ can be written as
functions of
$e (m_Z)$ and $\sin^2 \theta^\ell_{\mbox{eff}}$:
$$
g(m_Z) = {e(m_Z)\over \sin \theta^\ell_{\mbox{eff}}},
\eqno{\mbox{(2a)}}
$$
$$
g'(m_Z) = {e(m_Z)\over \cos \theta^\ell_{\mbox{eff}}},
\eqno{\mbox{(2b)}}
$$
$$
g_Z(m_Z) = {e(m_Z)\over \sin \theta^\ell_{\mbox{eff}}\cos
\theta^\ell_{\mbox{eff}}},
\eqno{\mbox{(2c)}}
$$
where $\theta^\ell_{\mbox{eff}}$ is the effective electroweak mixing
angle extracted from $Z$ leptonic decays \cite{1}.

In this paper we propose that the couplings $g, g', g_Z$ and $e$, at
the $Z$ mass scale, are functions on $\theta^\ell_{\mbox{eff}}$
only. Using the following ansatz
$$
g(m_Z) g'(m_Z) = g_Z(m_Z) \ e (m_Z) = \sin^2
\theta^\ell_{\mbox{eff}}\ ,
\eqno{(3)}
$$
we get from Eqs (2a-c) 
$$
g^2 (m_Z) = \sin \theta^\ell_{\mbox{eff}} \ \cos
\theta^\ell_{\mbox{eff}}\ , \hspace{18mm}
\eqno{\mbox{(4a)}}
$$
$$
g^2_Z (m_Z) = \tan \theta^\ell_{\mbox{eff}}\ , \hspace{31mm}
\eqno{\mbox{(4b)}}
$$
$$
g'^2 (m_Z) = \sin \theta^\ell_{\mbox{eff}} \ \cos
\theta^\ell_{\mbox{eff}} \ \tan^2 \theta^\ell_{\mbox{eff}}\ ,
\eqno{\mbox{(4c)}}
$$
$$
e^2 (m_Z) = \sin^3 \theta^\ell_{\mbox{eff}}  \ \cos
\theta^\ell_{\mbox{eff}}\ . \hspace{15mm}
\eqno{\mbox{(4d)}}
$$
Using Eq.(4d) and the result \cite{2}
$$
\alpha^{-1} (m_Z) = 128.952 \pm 0.049,
\eqno{(5)}
$$
we obtain
$$
\sin^2 \theta^\ell_{\mbox{eff}} = 0.23116 \pm 0.00007,
\eqno{(6)}
$$
to be compared with the experimental value \cite{1},
$$
\sin^2 \theta^\ell_{\mbox{eff}} = 0.23113 \pm 0.00021
\eqno{(7)}
$$
extracted from $Z$ leptonic decay data. We don't consider
$\sin^2 \theta^\ell_{\mbox{eff}}$ extracted from $Z$ hadronic
decay data \cite{1} because of possible hadronic complications.
 The comparison between Eqs (6) and (7) is impressive.
This is the main result of our paper.

Now, we define the $Z$ boson mass as
$$
\overline{m}_Z = g_Z (m_Z) {v\over 2}\ .
\eqno{(8)}
$$
Using Eqs (4b), (6) and
$$
v = \left({1\over \sqrt{2} G_F}\right)^{1/2} = 246.218(1)\
\mbox{GeV},
\eqno{(9)}
$$
where $G_F$ is the Fermi coupling constant \cite{3}, we get
$$
\overline{m}_Z = 91.1611 \pm 0.0086  \
\mbox{GeV}. 
\eqno{(10)}
$$
In order to compare this prediction of our ansatz with experiments, let
us remember that we can define the $Z$ boson mass at least in three
different ways: $m_Z$, $m_1$ and $m_2$ \cite{3}.
We proceed to describe the relations among these mass parameters.
We start from the $Z$ boson propagator proportional to 
$$
D^{-1} (s) = s- m^2_Z + i {s\over m_Z} \Gamma_Z \hspace{37mm}
\eqno{\mbox{(11a)}}
$$
$$
= \left[1+{\Gamma_1\over m_1}\right]  [s - m^2_1 + i \Gamma_1
m_1]\hspace{6mm}
\eqno{\mbox{(11b)}}
$$
$$
= \left[1+{\Gamma_1\over m_1}\right] \left[s - \left(m_2 - i
{\Gamma_2 \over  m_2}\right)^2\right]\ .
\eqno{\mbox{(11c)}}
$$
$m_Z$ and $\Gamma_Z$ are the usual $Z$ boson mass and width used
by experimentalists \cite{3}:
$$
m_Z = 91.1876 \pm 0.0021 \ \mbox{GeV},
\eqno{\mbox{(12a)}}
$$
$$
\Gamma_Z = 2.4952 \pm 0.0023 \ \mbox{GeV}.
\eqno{\mbox{(12b)}}
$$
To get the values of $(m_1, \Gamma_1)$ and $(m_2, \Gamma_2)$ defined at
the $Z$
pole, we use the following formulae derived from Eqs (11a-c):
$$
{\Gamma_Z \over m_Z} = {\Gamma_1\over m_1}, \hspace{42mm}
\eqno{\mbox{(13a)}}
$$
$$
\Gamma_1 m_1 = \Gamma_2 m_2, \hspace{41mm}
\eqno{\mbox{(13b)}}
$$
$$
m^2_1 = {m^2_Z \over 1+ {\Gamma^2_Z \over m^2_Z}}, \hspace{33mm}
\eqno{\mbox{(13c)}}
$$
$$
m^2_2 = {m^2_Z\over 1 + {\Gamma^2_Z\over m^2_Z}}
\left({1+\sqrt{1+{\Gamma^2_Z\over m^2_Z}}\over 2}\right).
\eqno{\mbox{(13d)}}
$$
With the help of Eqs (12a-b), (13c) and (13d), we obtain
$$
m_1 = 91.1535 \pm 0.0021 \ \mbox{GeV},
\eqno{\mbox{(14a)}}
$$
$$
m_2 = 91.1620 \pm 0.0021  \ \mbox{GeV}.
\eqno{\mbox{(14b)}}
$$
A comparison between Eqs (10), (12a), (14a) and (14b) indicates
that the central value in Eq.(10) is in very good agreement with
the one of Eq.(14b). Therefore, we propose
$$
m_2 = \overline{m}_Z = g_Z (m_Z) {v\over 2}.
\eqno{(15)}
$$
It is interesting to note that if we use Eqs (4b), (14b) and
(15), we get
$$
\alpha^{-1} (m_Z) = 128.946 \pm 0.011,
\eqno{(16)}
$$
$$
\sin^2 \theta^\ell_{\mbox{eff}} = 0.23117 \pm 0.00002,
\eqno{(17)}
$$
to be compared with Eqs (5) and (7).

\subsection*{Conclusions.}

From our ansatz given in Eqs (3) or (4d), we get Eq.(6) from
Eq.(5). This is the main result of our paper. Assuming Eq.(15), we obtain
Eq.(16) and (17). Agreement with experimental data (Eqs (5) and
(7)) is remarkable. It is perhaps fortuitous. If not, the
question is: why ?
\subsection*{Acknowledgments}
I would like to thank the organizers of this meeting for the invitation
and hospitality. I thank Gabriel L\'opez Castro for years of long
exchanges of view on the subject.

\bibliographystyle{aipproc}

\end{document}